\documentclass[aps,nofootinbib,showpacs,showkeys,twocolumn,longbibliography,prd,10pt]{revtex4-1}
\usepackage{xcolor}
\usepackage{amsmath,amssymb,array}
\usepackage[english]{babel}
\usepackage{tensor}
\usepackage{hyperref}
\hypersetup{
pdftitle={},%
pdfauthor={},%
pdfsubject={},%
pdfkeywords={},%
colorlinks=true,%
linkcolor=blue,%
citecolor=crimson,%
linktocpage=true,%
pageanchor=true,%
urlcolor=black
}
\definecolor{crimson}{rgb}{0.7, 0.08, 0.24}
\newcommand{\dd}{\mathrm{d}}
\newcommand{\partdif}[2]{\frac{\partial #1}{\partial #2}}
\newcommand{\ppartdif}[3]{\frac{\partial^2 #1}{\partial #2 \partial #3}}

\newcommand{\F}[0]{\mathcal{F}}
\renewcommand{\H}[0]{\mathcal{H}}
\renewcommand{\L}[0]{\mathcal{L}}
\newcommand{\xX}[2]{\partdif{x^{#1}}{X^{#2}}}
\newcommand{\Xx}[2]{\partdif{X^{#1}}{x^{#2}}}
\newcommand{\aphi}[0]{\bar{\phi}}
\newcommand{\aPhi}[0]{\bar{\Phi}}
\newcommand{\api}[0]{\bar{\pi}}
\newcommand{\aPi}[0]{\bar{\Pi}}
\newcommand{\ind}[1]{\indices{#1}}
\newcommand{\sla}[2]{\left. #1 \right|_{#2}}

\begin{document}
\title{Vanishing torsion coupling of the Maxwell field in canonical gauge theory of gravity}
\author{Johannes~M\"unch}
\email{johannes.muench@physik.uni-regensburg.de}
\affiliation{University of Regensburg, Universit\"atstrasse 31, 93040 Regensburg, Germany}
\author{J\"urgen~Struckmeier}
\email{struckmeier@fias.uni-frankfurt.de}
\affiliation{Frankfurt Institute for Advanced Studies (FIAS), Ruth-Moufang-Strasse~1, D-60438 Frankfurt am Main}
\affiliation{Goethe Universit\"at, Max-von-Laue-Strasse~1, D-60438~Frankfurt am Main}
\author{David~Vasak}
\email{vasak@fias.uni-frankfurt.de}
\affiliation{Frankfurt Institute for Advanced Studies (FIAS), Ruth-Moufang-Strasse~1, D-60438 Frankfurt am Main}
\affiliation{Goethe Universit\"at, Max-von-Laue-Strasse~1, D-60438~Frankfurt am Main}
\received{\today}

\begin{abstract}
The Maxwell field can be viewed as a $U(1)$-gauge theory, therefore, generalizing it to form-invariance
in dynamical spacetime backgrounds should take this symmetry into account.
This is of essential importance when generalizations of general relativity to theories with non-vanishing torsion are considered.
Despite the many statements in literature that a $U(1)$-gauge field cannot couple to torsion, this issue was recently revived.
In this letter we contribute to the discussion by demonstrating via a canonical transformation
within the framework of the DeDonder-Weyl Hamiltonian formalism that a $U(1)$-gauge field does \emph{not} couple to torsion.
\end{abstract}
\maketitle
\section{Introduction}
The generalization of electrodynamics to arbitrary spacetime backgrounds, hence its coupling to Einstein's general relativity is well-known.
The issue becomes more subtle when generalizations of general relativity, including non-vanishing torsion or non-metricity, are included.
Examples of these theories are Einstein-Cartan or teleparallel gravity.
In literature it was discussed in detail that conservation of electromagnetic current prohibits any couplings between the
electromagnetic gauge field $a_\mu$ and spacetime torsion $s\ind{^\alpha_{\mu \nu}} = \gamma\ind{^\alpha_{[\mu \nu]}}$
(see~\cite{HehlHowDoestheElectromagnetic,PuntigamMaxwellstheoryon,RubilarTorsionnonminimallycoupled}).

In recent works this topic gained again attention in the context of physical consequences of torsion coupling breaking $U(1)$-symmetry~\cite{CabralEinstein-Cartan-Dirac,CabralColsmologicalbouncescyclic, SakethTorsiondrivenInflationary}.

The DeDonder-Weyl (DW) Hamiltonian formalism is a covariant formalism, i.e.\ spatial and timelike directions are treated on equal footing,
which allows to deploy the covariant canonical transformation theory in field theories for implementing a local symmetry described by arbitrary Lie groups.
Therein, generic transformations such as gauge transformations and in particular spacetime diffeomorphisms can be formulated in the language
of generating functionals (see~\cite{StruckmeierCanonicalTransformationPath,StruckmeierCanonicaltransforamtionpathII}).
Its specific advantage is that there is no more input needed beyond the transformation behavior of the initial fields to systematically
derive the kind of gauge fields, their transformation behavior, and their coupling to the initial fields in order to render the theory invariant under this transformation.
The reason for this is that a generating functional simultaneously specifies the transformation rules for the field \emph{and} their
canonical conjugate fields, in conjunction with the appropriate transformation rule for the Hamiltonian.
In this letter, the formalism is applied to $U(1)\times\text{Diff}(M)$ gauge transformations to determine the couplings between a $U(1)$-gauge
field $a_\mu$ and spacetime torsion $s\ind{^\alpha_{\mu \nu}}$.
This strategy focuses on the $U(1)$-symmetry of electrodynamics and contributes to the discussion of
\cite{HehlHowDoestheElectromagnetic,PuntigamMaxwellstheoryon,RubilarTorsionnonminimallycoupled} as an independent approach.

The paper is structured as follows:
In Sec.~\ref{sec:CovdeDonder} the DW Hamiltonian formalism is reviewed at the example of the Maxwell theory.
In Sec.~\ref{sec:U1torsion}, the couplings between $U(1)$-gauge fields and torsion are systematically derived.
The paper closes with the conclusions in Sec.~\ref{sec:conclusions}.
\section{Covariant DeDonder-Weyl Hamiltonian Formalism}\label{sec:CovdeDonder}
We first review the covariant DW Hamiltonian formalism \cite{dedonder30,weyl35} at the example of
a real-valued scalar field theory $\tilde{\L}_M(\phi,\partial\phi, g)$ that couples to a gravitational theory,
specified by $\tilde{\L}_{grav}(g,\partial g ,\gamma, \partial \gamma)$.
The starting point is the action
\begin{align}
&\quad\, S(\phi,\partial \phi,g_{\mu \nu}, \partial g_{\mu \nu}, \gamma\ind{^\alpha_{\mu \nu}}, \partial \gamma\ind{^\alpha_{\mu \nu}})
\label{eq:initial-action}\\
&=\int\dd^4 x\left[ \tilde{\L}_M(\phi,\partial\phi, g) + \tilde{\L}_{grav}(g,\partial g ,\gamma, \partial \gamma)+\tilde{\L}_{g}\right],
\notag
\end{align}
where $g_{\mu \nu}(x)$ is the spacetime metric and $\gamma\ind{^\alpha_{\mu \nu}}(x)$ an arbitrary spacetime connection.
In order for the action~\eqref{eq:initial-action} to be diffeomorphism-invariant, the total Lagrangian
$\tilde{\L}=\tilde{\L}_M + \tilde{\L}_{grav}+\tilde{\L}_{g}$ must be a world scalar density, hence a relative scalar of weight $w=1$, denoted by the tilde.
The matter Lagrangian $\tilde{\L}_M$ represents an arbitrary Lorentz-invariant Lagrangian density on Minkowski spacetime,
which is trivially generalized to a dynamic spacetime by replacing the metric $\eta_{\mu \nu} \mapsto g_{\mu \nu}$.
The gravity Lagrangian $\tilde{\L}_{grav}$ stands for the Einstein-Hilbert Lagrangian in the Palatini formulation
or for any other theory of the ``free'' (uncoupled) gravitational field, e.g.\ a higher-curvature gravity theory.
Finally, the gauge Lagrangian $\tilde{\L}_{g}$ is to be constructed to make the Lagrangian $\tilde{\L}_M+\tilde{\L}_{grav}$
into a world scalar density $\tilde{\L}$ in order for the action~\eqref{eq:initial-action} to be diffeomorphism-invariant.
The DW poly-momenta are defined \emph{locally} as
\begin{align}\label{eq:polymomenta}
\tilde{\pi}^{\mu} &= \partdif{\tilde{\L}}{\left(\partial_\mu \phi\right)} \;,& \tilde{k}^{\mu \nu \alpha} &=
\partdif{\tilde{\L}}{\left(\partial_\alpha g_{\mu \nu}\right)}\;,
\notag\\
\tilde{q}\ind{_\alpha^{\mu \nu \beta}} &=  \partdif{\tilde{\L}}{\left(\partial_\beta \gamma\ind{^\alpha_{\mu \nu}}\right)}\;,
\end{align}
which must all turn out to be proper tensor densities once the gauge terms have been added to the then gauge-invariant Lagrangian $\tilde{\L}$.
In contrast to the conventional Hamiltonian formulation, the DW formalism assigns four momenta to each field component,
which reflects the duality of moments and derivatives (timelike and spatial).

The DeDonder-Weyl Hamiltonian is then defined via the Legendre-transformation
\begin{equation}\label{eq:legendre-def}
\tilde{\H}(\phi,\pi,g,k,\gamma,q) \!=\! \tilde{\pi}^\mu \partial_\mu \phi +\tilde{k}^{\mu \nu \alpha} \partial_\alpha g_{\mu \nu}
+ \tilde{q}\ind{_\alpha^{\mu \nu \beta}} \partial_\beta \gamma\ind{^\alpha_{\mu \nu}} - \tilde{\L}.
\end{equation}
With Eq.~\eqref{eq:polymomenta} and assuming regularity of the Legendre transformation~\eqref{eq:legendre-def},
it is possible to replace all derivatives by means of the poly-momenta which yields the DW Hamiltonian density $\tilde{\H}$.
In the case of the Hilbert Lagrangian with its \emph{linear} dependence of the Riemann tensor, this not possible and leads to
primary constraints according to Dirac's constraint theory~\cite{DiracLecturesOnQuantum} (see also~\cite{DateLecturesonConstrained}).
Commonly, symmetry constraints, such as $\tilde{k}^{[\mu \nu] \alpha} = 0$, are present in this formalism, which need to be taken into account.
The dynamics of the system is then described by the DW equations
\footnote{The above mentioned symmetry constraints can be considered in the variation of the action. This leads to symmetrized DeDonder-Weyl canonical equations.}
\begin{align}
\partial_\mu \phi &= \partdif{\tilde{\H}}{\tilde{\pi}^{\mu}}\;,&
\partial_\mu\tilde{\pi}^{\mu} &= -\partdif{\tilde{\H}}{\phi}\notag\\
\partial_\alpha g_{\mu \nu} &= \partdif{\tilde{\H}}{\tilde{k}^{\mu \nu \alpha }} \;,&
\partial_\alpha \tilde{k}^{\mu \nu \alpha} &= -\partdif{\tilde{\H}}{g_{\mu \nu}}\notag\\
\partial_\beta \gamma\ind{^\alpha_{\mu \nu}} &= \partdif{\tilde{\H}}{\tilde{q}\ind{_\alpha^{\mu \nu \beta}}} \,,&
\partial_\beta \tilde{q}\ind{_\alpha^{\mu \nu \beta}} &= -\partdif{\tilde{\H}}{\gamma\ind{^\alpha_{\mu \nu}}}
\end{align}
and is equivalent to the conventional Hamiltonian formulation, but covariant, i.e.\ without the need for a $3+1$-split.

Beside the unified treatment of space and time, the DW formulation has the advantage that the whole machinery of
generating functionals for canonical transformations exists~\cite{StruckmeierCanonicalTransformationPath}.
In the actual context, useful generating functionals are $\tilde{\F}_2^\mu(\phi,\tilde{\Pi},g,\tilde{K}, \gamma,\tilde{Q})$
of type $\tilde{\F}_2$ with the transformation rules
\allowdisplaybreaks
\begin{align}
\delta_{\nu}^{\mu} \Phi(X) =& \; \partdif{\tilde{\mathcal{F}}_2^{\kappa}}{\tilde{\Pi}^{\nu}}
\partdif{X^{\mu}}{x^{\kappa}} \left|\xX{}{}\right| \;,\\
\tilde{\pi}^{\mu} (x) =& \; \partdif{\tilde{\mathcal{F}}_2^{\mu}}{\phi} \;,
\label{eq:trafoPgeneric}\\
\delta_{\nu}^{\mu} G_{\alpha \lambda}(X) =& \; \partdif{\tilde{\F}_2^{\kappa}}{\tilde{K}^{\alpha \lambda \nu}} \partdif{X^{\mu}}{x^{\kappa}} \left|\xX{}{}\right|\\
\tilde{k}^{\alpha \lambda \mu}(x) =& \; \partdif{\tilde{\F}_2^{\mu}}{g_{\alpha \lambda}} \;,
\label{eq:trafokgeneric}\\
\delta_{\nu}^{\mu} \Gamma\ind{^\beta_{\alpha\lambda}}(X) =& \; \partdif{\tilde{\F}_2^{\kappa}}{\tilde{Q}\ind{_\beta^{\alpha\lambda\nu}} }
\partdif{X^{\mu}}{x^{\kappa}} \left|\xX{}{}\right|\\
\tilde{q}\ind{_{\beta}^{\alpha\lambda\mu}} =& \; \partdif{\tilde{\mathcal{F}}_2^{\mu}}{\gamma\ind{^{\beta}_{\alpha\lambda}}} \;,
\label{eq:trafoQgeneric}\\
\sla{\tilde{\mathcal{H}}'}{X} =& \; \left( \sla{\tilde{\mathcal{H}}}{x} +
\left. \partdif{\tilde{\F}_2^{\alpha}}{x^{\alpha}}\right|_{expl.} \right) \left|\xX{}{}\right| \;,
\label{eq:Hamiltongeneric}
\end{align}
where upper case letters denote the transformed quantities.
In this description also coordinate transformations \mbox{$x \mapsto X$} are permitted, with $\left|\partial x/\partial X\right|$ denoting its Jacobi-determinant.

As shown in \cite{StruckmeierCanonicalTransformationPath,StruckmeierCanonicaltransforamtionpathII},
gauge transformations as $SU(N)$ or $\text{Diff}(M)$ can be formulated in terms of canonical transformations generated by $\tilde{\F}_2^\mu$.
This way the necessity for introducing a gauge field to render the system gauge invariant,
its transformation behavior and the couplings between initial field and gauge field
can systematically be derived without any further assumptions.
In the following, this formalism is used to demonstrate that no coupling between $U(1)$-gauge fields $a_\mu$ and a
spacetime torsion $s\ind{^\alpha_{\mu \nu}} = \gamma\ind{^\alpha_{[\mu \nu]}}$ exists that maintains the system's $U(1)\times\text{Diff}(M)$ invariance.
\section{Coupling of Maxwell-Field to Torsion}\label{sec:U1torsion}
We use the formalism of covariant canonical transformations and apply them to a system of a complex scalar field $\phi$
and a $U(1)$-gauge field $a_\mu$ which are coupled to Palatini gravity, the latter represented by both the metric
$g_{\mu \nu}$ and the spacetime connection $\gamma\ind{^\alpha_{\mu \nu}}$ as \emph{a priori} independent fields.
Note that the spacetime connection $\gamma\ind{^\alpha_{\mu \nu}}$ is arbitrary, hence possibly comprises torsion
and non-metricity, i.e., a not covariantly conserved metric.
Consider a Hamiltonian system $\tilde{\H}$ consisting of the three DW Hamiltonian densities
\begin{align*}
\tilde{\H} =& \tilde{\H}_1(\phi,\aphi,\tilde{\pi}^\mu,\tilde{\api}^\mu;g_{\mu \nu}) + \tilde{\H}_2(a_\mu,p^{\mu \nu};g_{\mu \nu})\\
&\;+ \tilde{\H}_{grav}(g_{\mu\nu}, \tilde{k}^{\mu \nu \alpha}, \gamma\ind{^\alpha_{\mu \nu}}, \tilde{q}\ind{_\alpha^{\mu \nu \rho}})+ \tilde{\H}_{g} \;,
\end{align*}
where the fields $(\aphi,\tilde{\pi}^\mu)$, $(\phi,\tilde{\api}^\mu)$, $(a_\mu,\tilde{p}^{\mu \nu})$, $(g_{\mu\nu}, \tilde{k}^{\mu \nu \alpha})$,
and $(\gamma\ind{^\alpha_{\mu \nu}}, \tilde{q}\ind{_\alpha^{\mu \nu \rho}})$ constitute canonical pairs.
The Hamiltonian is considered invariant under \textit{global} $U(1)\times\text{Diff}(M)$ ($\Lambda = const.$, $\partial X/ \partial x = const.$) transformations.
The aim is to construct the gauge Hamiltonian $\tilde{\H}_{g}$ that renders the total system
$\tilde{\H}$ invariant under both local $U(1)$-transformations as well as generic diffeomorphisms
$f \in \text{Diff}(M)$.
The canonical transformation formalism allows to derive the interaction terms which make the system gauge invariant.
Here we already included the required gauge fields $a_\mu$ and $\gamma\ind{^\alpha_{\mu \nu}}$.
These could also be introduced via a step-by-step derivation (see~\cite{StruckmeierCanonicalTransformationPath}), which need not be repeated here.
A generic transformation in $U(1)\times \text{Diff}(M)$ acts on the fields as
\allowdisplaybreaks
\begin{subequations}\label{eq:trafoall}
\begin{align}
\phi(x) \mapsto \Phi(X) &= \phi(x) e^{i\Lambda(x)}\;,\label{eq:trafophi}\\
\aphi(x) \mapsto \aPhi(X) &= \aphi(x) e^{-i\Lambda(x)}\\
a_{\mu}(x) \mapsto A_{\mu}(X) &= \left(a_\xi + i \partdif{\Lambda(x)}{x^\xi}\right) \xX{\xi}{\mu}\;,\label{eq:trafoa}
\end{align}
\begin{align}
g_{\mu \nu}(x) \mapsto G_{\mu \nu}(X) &= g_{\sigma \rho}(x) \xX{\sigma}{\mu} \xX{\rho}{\nu} \;,\label{eq:trafog}\\
\gamma\ind{^\alpha_{\mu \nu}}(x) \mapsto \Gamma\ind{^\alpha_{\mu \nu}}(X)
&= \gamma\ind{^\beta_{\sigma \rho}}(x) \Xx{\alpha}{\beta} \xX{\sigma}{\mu} \xX{\rho}{\nu} \label{eq:trafogamma}\notag\\
&\quad+ \Xx{\alpha}{\xi} \frac{\partial^2 x^{\xi}}{\partial X^{\mu} \partial X^{\nu}} \;.
\end{align}
\end{subequations}
We abbreviate the map $X\circ f \circ x^{-1}$, where $x$ is a chart in $U \subset M$ and $X$ a chart in $f(U) \subset M$ simply as $X$.
The metric transforms then according to the pullback along $f^{-1}$, where $\partial x^\sigma/\partial X^\mu = \partial_\mu(x\circ f^{-1} \circ X)^\sigma$
and similar $\partial X^\sigma/\partial x^\mu= \partial_\mu(X\circ f \circ x)^\sigma$.
The $U(1)$-connection $a_\mu$ is first transformed by a local $U(1)$-transformation and then also pulled back along $f^{-1}$
and $\gamma\ind{^\alpha_{\mu \nu}}$ has the usual inhomogeneous transformation rule for spacetime connections.
\begin{widetext}
The generating functional which generates the transformations~\eqref{eq:trafoall} is given by
\begin{align}
\tilde{\mathcal{F}}_2^{\mu} =& \; \left[ \tilde{\aPi}^{\beta}(X) \phi(x) e^{i\Lambda(x)} + \aphi(x) \tilde{\Pi}^{\beta}(X)
e^{-i\Lambda(x)} + \tilde{P}^{\alpha \beta}(X)\left( a_{\xi}(x) + i \partdif{\Lambda(x)}{x^{\xi}} \right) \frac{\partial x^{\xi}}{\partial X^{\alpha}} \right.
\notag\\
& \; + \left. \tilde{K}^{\eta \xi \beta}(X) g_{\sigma \rho}(x) \xX{\sigma}{\eta} \xX{\rho}{\xi}
+ \tilde{Q}\ind{_{\lambda}^{\alpha \rho \beta}}(X) \left( \gamma\ind{^{\sigma}_{\xi \eta}}(x) \partdif{x^{\xi}}{X^{\alpha}}
\partdif{x^{\eta}}{X^{\rho}} \partdif{X^{\lambda}}{x^{\sigma}} + \frac{\partial^2 x^{\xi}}{\partial X^{\alpha} \partial X^{\rho}}
\partdif{X^{\lambda}}{x^{\xi}} \right)\right] \frac{\partial x^{\mu}}{\partial X^{\beta}} \left|\xX{}{}\right|^{-1} \;.
\end{align}
\end{widetext}
In addition to Eqs.~\eqref{eq:trafophi}--\eqref{eq:trafogamma}, the generating functional also determines the transformations
for the canonical momenta and for the Hamiltonian
\allowdisplaybreaks
\begin{subequations}\label{eq:mtrafoall}
\begin{align}
\tilde{\pi}^{\mu}(x) =& \; \frac{ \partial \tilde{\mathcal{F}}_2^{\mu}}{\partial \aphi}
= \tilde{\Pi}^{\beta}(X) e^{-i\Lambda(x)} \partdif{x^{\mu}}{X^{\beta}}  \left|\xX{}{}\right|^{-1} \;,
\label{eq:trafopi}\\
\tilde{\bar{\pi}}^{\mu}(x) =& \; \frac{ \partial \tilde{\mathcal{F}}_2^{\mu}}{\partial \phi}
= \tilde{\bar{\Pi}}^{\beta}(X) e^{i\Lambda(x)} \partdif{x^{\mu}}{X^{\beta}} \left|\xX{}{}\right|^{-1} \;,
\label{eq:trafoapi}\\
\tilde{p}^{\mu \nu} (x) =& \; \partdif{\tilde{\mathcal{F}}_2^{\nu}}{a_{\mu}}
= \tilde{P}(X)^{\alpha \beta} \partdif{x^{\mu}}{X^{\alpha}} \partdif{x^{\nu}}{X^{\beta}} \left|\xX{}{}\right|^{-1} \;,
\label{eq:trafoP}\\
\tilde{k}^{\alpha \lambda \mu}(x) =& \; \partdif{\tilde{\F}_2^{\mu}}{g_{\alpha \lambda}}
= \tilde{K}^{\eta \xi \beta}(X) \xX{\alpha}{\eta} \xX{\lambda}{\xi} \xX{\mu}{\beta} \left|\xX{}{}\right|^{-1} \;,
\label{eq:trafok}\\
\tilde{q}\ind{_{\beta}^{\epsilon \zeta \mu}} =& \; \partdif{\tilde{\mathcal{F}}_2^{\mu}}{b\ind{^{\beta}_{\epsilon \zeta}}}
= \tilde{Q}\ind{_{\lambda}^{\alpha \rho \kappa}} \partdif{X^{\lambda}}{x^{\beta}} \partdif{x^{\epsilon}}{X^{\alpha}}
\partdif{x^{\zeta}}{X^{\rho}} \partdif{x^{\mu}}{X^{\kappa}} \!\left| \xX{}{} \right|^{-1}\!\!\!\!,
\label{eq:trafoQ}\\
\sla{\tilde{\mathcal{H}}^{\prime}}{X} =& \; \left( \sla{\tilde{\mathcal{H}}}{x}
+ \left. \partdif{\tilde{\F}_2^{\alpha}}{x^{\alpha}}\right|_{expl.} \right) \left|\xX{}{}\right|\;.
\label{eq:Hamilton}
\end{align}
\end{subequations}
The relevant gauge couplings are obtained from the transformation rule~\eqref{eq:Hamilton} for the Hamiltonian.
Using the identity $\partdif{}{x^\mu}\left(\frac{\partial x^{\mu}}{\partial X^{\beta}} \left|\xX{}{}\right|^{-1}\right) \equiv 0$, one finds
\allowdisplaybreaks
\begin{align}
&\left. \partdif{\tilde{\F}_2^{\alpha}}{x^{\alpha}}\right|_{expl.} \left|\xX{}{}\right|
= \left[\left(\tilde{\bar{\Pi}}^\beta \phi e^{i\Lambda} - \aphi \tilde{\aPi}^\beta e^{-i\Lambda}\right) i \partdif{\Lambda}{x^\mu} \right.
\label{eq:trafoU1}\\
&\left.+\, \tilde{P}^{\alpha \beta} \left(a_\xi + i \partdif{\Lambda}{x^\xi}\right) \ppartdif{x^\xi}{X^\alpha}{X^\eta} \Xx{\eta}{\mu} \right.
\label{eq:Pgauge}\\
&\left.+\, \tilde{P}^{\alpha\beta} i \ppartdif{\Lambda}{x^\xi}{x^\mu} \xX{\xi}{\alpha}
+ (\tilde{K}, g, \tilde{Q},\gamma)\right] \xX{\mu}{\beta} \;,
\label{eq:Pgamma}
\end{align}
where the abbreviation $ (\tilde{K}, g, \tilde{Q},\gamma)$ stands for terms which are discussed in detail
in~\cite{StruckmeierCanonicalTransformationPath,StruckmeierCanonicaltransforamtionpathII} and are not relevant in the actual context.
Inserting now the transformation rules~\eqref{eq:trafophi}-\eqref{eq:trafogamma} and~\eqref{eq:trafopi}--\eqref{eq:trafoQ}
provides the gauge couplings which make the system invariant under the transformation group $U(1)\times \text{Diff}(M)$.
Especially important are the rules~\eqref{eq:trafoa} and~\eqref{eq:trafogamma}, which can be used to eliminate the respective inhomogeneous terms
\begin{align}
i\partdif{\Lambda}{x^\mu} &= \Xx{\eta}{\mu} A_\eta - a_\mu \;, \label{eq:lambdaDV}\\
\ppartdif{x^\xi}{X^\alpha}{X^\eta} \Xx{\eta}{\mu} &= \Gamma\ind{^\xi_{\alpha \mu}}
- \gamma\ind{^\eta_{\beta \lambda}} \xX{\beta}{\alpha} \xX{\lambda}{\mu} \Xx{\xi}{\eta} \;.
\label{eq:gamma2}
\end{align}
These replacements lead to the couplings of fields and gauge fields, which yield in particular for the term~\eqref{eq:trafoU1}:
\begin{align*}
&\quad\,\left(\tilde{\bar{\Pi}}^\beta \phi e^{i\Lambda} - \aphi \tilde{\aPi}^\beta e^{-i\Lambda}\right) i \partdif{\Lambda}{x^\mu}\xX{\mu}{\beta}\\
&=\left(\tilde{\aPi}^\mu\Phi - \aPhi\tilde{\Pi}^\mu\right) A_\mu - \left(\tilde{\api}^\mu\phi - \aphi\tilde{\pi}^\mu\right) a_\mu \left|\xX{}{}\right|,
\end{align*}
whereas for~\eqref{eq:Pgauge} and~\eqref{eq:Pgamma}:
\begin{align}
&\quad\,\tilde{P}^{\alpha \beta}\left[ \left(a_\xi + i \partdif{\Lambda}{x^\xi}\right) \ppartdif{x^\xi}{X^\alpha}{X^\beta}+
i \ppartdif{\Lambda}{x^\xi}{x^\mu} \xX{\xi}{\alpha}\xX{\mu}{\beta}\right]\notag\\
&=\tilde{P}^{(\alpha \beta)}\partdif{A_\beta}{X^\alpha}-\tilde{p}^{(\xi\mu)}\partdif{a_\xi}{x^\mu}\left|\xX{}{}\right|\label{eq:term2}\\
&\quad+\tilde{P}^{(\alpha \beta)}A_\rho\left( \ppartdif{x^\xi}{X^\alpha}{X^\beta}\Xx{\rho}{\xi} +
\ppartdif{X^\rho}{x^\xi}{x^\mu}\xX{\xi}{\alpha}\xX{\mu}{\beta}\right).\notag
\end{align}
By virtue of the identity
\begin{equation}\label{Xinv}
\ppartdif{x^\xi}{X^\alpha}{X^\beta} \partdif{X^{\rho}}{x^{\xi}}
\equiv-\ppartdif{X^\rho}{x^\xi}{x^\mu} \partdif{x^{\xi}}{X^{\alpha}}\partdif{x^{\mu}}{X^{\beta}}
\end{equation}
the last line of~\eqref{eq:term2} vanishes identically.
Note that the cancelling occurs due to the second $\Lambda$-derivatives in Eq.~\eqref{eq:Pgamma},
which are merely present in the case where a $U(1)$-invariance is demanded \emph{in addition} to the $\text{Diff}(M)$-invariance.
In the case of pure $\text{Diff}(M)$ transformations, the second derivative term of $\Lambda$ in Eq.~\eqref{eq:Pgamma} does not occur
and hence cannot cancel the second derivative term of $x^\xi$ in~Eq.~\eqref{eq:Pgauge}.
That ultimately leads to couplings of the form $\tilde{p}^{\mu \nu} a_\alpha \gamma\ind{^\alpha_{\mu \nu}}$
which in the final action functional convert the partial derivatives of $a_\mu$ into covariant derivatives.
In total the transformation rule~\eqref{eq:Hamilton} of the Hamiltonian becomes
\allowdisplaybreaks
\begin{align*}
&\left.\partdif{\tilde{\F}_2^{\alpha}}{x^{\alpha}}\right|_{expl.}\!\!
=\left( \tilde{\aPi}^{\alpha} \Phi + \aPhi \tilde{\Pi}^{\alpha} \right) A_\alpha\left|\Xx{}{}\right|
-\left( \tilde{\api}^{\alpha} \phi + \aphi \tilde{\pi}^{\alpha} \right) a_\alpha\\
&+\frac{1}{2} \tilde{P}^{\nu \beta} \left( \partdif{A_{\nu}}{X^{\beta}} + \partdif{A_{\beta}}{X^{\nu}} \right) \left|\Xx{}{}\right|
 -\frac{1}{2} \tilde{p}^{\nu \beta} \left( \partdif{a_{\nu}}{x^{\beta}} + \partdif{a_{\beta}}{x^{\nu}} \right)\\
&+\left(\tilde{K}^{\alpha\lambda\beta}G_{\mu\lambda}+\tilde{K}^{\lambda\alpha\beta}G_{\lambda\mu}\right)\Gamma\ind{^\mu_{\beta\alpha}}\left|\Xx{}{}\right|\\
&-\left(\tilde{k}^{\alpha\lambda\beta}g_{\mu\lambda}+\tilde{k}^{\lambda\alpha\beta}g_{\lambda\mu}\right)\gamma\ind{^\mu_{\beta\alpha}}\\
&+(\tilde{Q}, \Gamma)\left|\Xx{}{}\right|-(\tilde{q}, \gamma)\\
&=\tilde{\H}_g^{\prime}\left|\Xx{}{}\right| - \tilde{\H}_g\;,
\end{align*}
which shows that the gauge Hamiltonian $\tilde{\H}_g$ emerges as:
\begin{align}
\tilde{\H}_g&=\left( \tilde{\api}^{\alpha} \phi + \aphi \tilde{\pi}^{\alpha} \right) a_\alpha+
\frac{1}{2} \tilde{p}^{\nu \beta} \left( \partdif{a_{\nu}}{x^{\beta}} + \partdif{a_{\beta}}{x^{\nu}} \right)\notag\\
&\quad+\left(\tilde{k}^{\alpha\lambda\beta}g_{\mu\lambda}+\tilde{k}^{\lambda\alpha\beta}g_{\lambda\mu}\right)\gamma\ind{^\mu_{\beta\alpha}}+(\tilde{q}, \gamma).
\end{align}
Again the explicit form of the $(\tilde{q}, \gamma)$ terms are not relevant in our context.
The gauge Hamiltonian $\tilde{\H}_g$ and $\tilde{\H}^{\prime}_g$ are form-invariant.
Hence, the total Hamiltonian $\tilde{\H}=\tilde{\H}_1+\tilde{\H}_2+\tilde{\H}_{grav}+\tilde{\H}_g$ transforms invariantly as a world scalar density.
The gauge Hamiltonian $\tilde{\H}_g$ does not induce couplings between vector field $a_\mu$ and the connection $\gamma\ind{^\alpha_{\mu \nu}}$---and hence torsion.
In contrast, the metric $g_{\mu \nu}$ couples directly to $\gamma\ind{^\alpha_{\mu \nu}}$, which is also the case
for non-$U(1)$-symmetric vector fields (cf.~\cite{StruckmeierCanonicalTransformationPath,StruckmeierCanonicaltransforamtionpathII}).

We finally end up with the $U(1)\times \text{Diff}(M)$-invariant action:
\begin{widetext}
\begin{align*}
S =& \int\!\dd^4 x\!\left[\tilde{\pi}^\mu \partial_\mu \aphi + \partial_\mu \phi \tilde{\api}^\mu
+ \tilde{p}^{\mu \nu} \partial_\nu a_\mu + \tilde{k}^{\alpha \beta \mu} \partial_\mu g_{\alpha \beta}
+ \tilde{q}\ind{_{\mu}^{\alpha \beta \nu}} \partial_\nu \gamma\ind{^\mu_{\alpha \beta}} - \tilde{\H}_1 - \tilde{\H}_2 - \tilde{\H}_{grav} - \tilde{\H}_g\right]\\
=&\int\!\dd^4 x\!\left[\tilde{\pi}^\mu \left(\partial_\mu \aphi - \aphi a_\mu\right) + \left(\partial_\mu \phi
- \phi a_\mu\right) \tilde{\api}^\mu + \frac{1}{2}\tilde{p}^{\mu \nu}\!\left(\partial_\nu a_\mu - \partial_\mu a_\nu\right)
+ \tilde{k}^{\alpha \beta \mu} \nabla_\mu g_{\alpha \beta} + \tilde{q}\ind{_{\mu}^{\alpha \beta \nu}} r\ind{^\mu_{\alpha \beta \nu}}
- \tilde{\H}_1 - \tilde{\H}_2 - \tilde{\H}_{grav}\right]
\end{align*}
\end{widetext}
where $\nabla_\mu$ is the spacetime-covariant derivative and $r\ind{^\mu_{\alpha \beta \nu}}$ the Riemann-Cartan curvature tensor.
It is easy to check that this action is indeed invariant under $U(1)\times\text{Diff}(M)$, as required.
The couplings of the kind $\tilde{\pi}^\mu \aphi\,a_\mu$ nicely add up to the $U(1)$-gauge-covariant derivative in the action.
On the other hand, merely the terms $\tilde{p}^{\mu \nu} \left(\partial_\nu a_\mu - \partial_\mu a_\nu\right)$ emerge,
yet no additional terms coupled to the affine connection arise.
Note that in~\cite{StruckmeierCanonicalTransformationPath}, where no $U(1)$-transformations were taken into account, this term looks like
\begin{equation*}
\tilde{p}^{\mu \nu} \left(\nabla_\nu a_\mu - \nabla_\mu a_\nu\right) = \tilde{p}^{\mu \nu} \left(\partial_\nu a_\mu
- \partial_\mu a_\nu\right) + \tilde{p}^{\mu \nu} a_\alpha s\ind{^\alpha_{\mu \nu}} \;,
\end{equation*}
which admits an explicit $U(1)$-breaking due to a coupling to torsion.

In the case of e.g.\ the Klein-Gordon-Maxwell system this gives the correct $U(1)$-gauge coupling and is diffemorphism invariant, as required.
It is obvious that the electromagnetic field strength tensor is given by
\begin{equation*}
F_{\mu \nu} = \partial_\mu a_\nu - \partial_\nu a_\mu \;,
\end{equation*}
in contrast to other proposals~\cite{CabralEinstein-Cartan-Dirac,CabralColsmologicalbouncescyclic}, which define
\begin{equation*}
\F_{\mu \nu} = \nabla_\mu a_\nu - \nabla_\nu a_\mu  = F_{\mu \nu} + a_\alpha s\ind{^\alpha_{\mu \nu}} \;,
\end{equation*}
and induce torsion couplings $a_\alpha s\ind{^\alpha_{\mu \nu}}$ explicitly breaking $U(1)$-symmetry.
\section{Conclusions}\label{sec:conclusions}
By means of a canonical transformation representation of the gauge principle, we again demonstrated that $U(1)$-symmetric vector fields do not couple minimally to torsion.
This conclusion coincides with~\cite{HehlHowDoestheElectromagnetic,PuntigamMaxwellstheoryon,RubilarTorsionnonminimallycoupled}
and supports their result with this independent account\footnote{Using the covariant transformation theory as described here this conclusion
can---in a straightforward albeit algebraically more elaborate way---be generalized to $SU(N)$ gauge theories~\cite{struckmeierprivatecommunication}.
This can be understood by noticing that the gauge connection $a = a_\mu \dd x^\mu$ is a Lie algebra valued 1-form. 
The corresponding field strength, i.e. curvature on the principal fibre bundle $F = \dd a + a \wedge a/2$, is then a Lie algebra valued 2-form.
This 2-form can be defined independently of any metric and connection and is thus covariant from the beginning.
It is not needed to add any further couplings --- as the formalism demonstrates.}.
Note that in principle non-minimal couplings to torsion, as considered in~\cite{RubilarTorsionnonminimallycoupled} of the form $s^2 F^2$, are still possible.
These couplings are $U(1) \times \text{Diff}(M)$ invariant and are of course not excluded by the covariant canonical transformation formalism.
\section*{Acknowledgments}
JM was supported by an International Junior Research Group grant of the Elite Network of Bavaria.
JS and DV thank the \emph{Walter Greiner-Gesellschaft zur F\"{o}rderung der physikalischen Grundlagenforschung e.V.},
and DV especially the \emph{Fueck-Foundation}, for support.

\end{document}